\documentclass[12pt]{article} 

\usepackage{amssymb} 
\usepackage{amsmath}
\usepackage{amsfonts}

\newcommand{\R}{\mathbb{R}}

\newcommand{\be}{\begin{equation}}
\newcommand{\bea}{\begin{eqnarray}}
\newcommand{\eea}{\end{eqnarray}}
\newcommand{\nn}{\nonumber}
\newcommand{\kt}{\rangle}
\newcommand{\br}{\langle}

\newcommand{\ed}{\end{document}}
\newcommand{\bbr}{\br\!\br}
\newcommand{\kkt}{\kt\!\kt}

\begin{document}

\title{Pseudo-Hermiticity and Generalized $PT$- and $CPT$-Symmetries}
\author{Ali Mostafazadeh\thanks{E-mail address: amostafazadeh@ku.edu.tr}\\ \\
Department of Mathematics, Ko\c{c} University,\\
Rumelifeneri Yolu, 80910 Sariyer, Istanbul, Turkey}
\date{ }
\maketitle

\begin{abstract}
We study certain linear and antilinear symmetry generators and involution operators associated with
pseudo-Hermitian Hamiltonians and show that the theory of pseudo-Hermitian operators provides a simple explanation for the recent results of Bender, Brody and Jones (quant-ph/0208076) on the $CPT$-symmetry of a class of $PT$-symmetric non-Hermitian Hamiltonians. We present a natural extension of these results to the class of diagonalizable pseudo-Hermitian Hamiltonians $H$ with a discrete spectrum. In particular, we introduce generalized parity (${\cal P}$), time-reversal (${\cal T}$), and charge-conjugation (${\cal C}$) operators and establish the 
${\cal PT}$- and ${\cal CPT}$-invariance of $H$.
\end{abstract}
%\vspace{2mm}
%PACS numbers: 03.65.Bz\\
%\vspace{2mm}

\baselineskip=24pt

\section{Introduction}

Bender, Brody and Jones \cite{bbj} have recently shown that for the class of $PT$-symmetric Hamiltonians
	\be
	H_\nu=p^2+x^2(ix)^\nu,~~~~~~~~\nu\in[0,\infty),
	\label{h}
	\end{equation}
one can use a complete set of eigenfunctions $\psi_n$ to construct a linear operator $C$ with the following properties.
	\begin{enumerate}
	\item $C$ is an involution generating a symmetry of the system, i.e., 
		\be
		C^2=1,~~~~~~~~~~~~~~[C,H_\nu]=0.
		\label{inv}
		\end{equation}
	In particular, $H_\nu$ is $CPT$-invariant;
	\item In the position representation, $C$ has the form
		\be
		C(x,y)=\sum_n \psi_n(x)\psi_n(y),~~~~~~~~\forall	x,y\in\R;
		\label{x-rep}
		\end{equation}
	\item The inner product
		\be
		\br\phi|\psi\kt_{CPT}:=\int_{\gamma} dx\, [CPT \phi(x)]\psi(x)
		\label{cpt-inner}
		\end{equation}
	is positive-definite, and the eigenfunctions $\psi_n$ are orthonormal with respect to this inner 	product, i.e.,
		\be
		\br\psi_m|\psi_n\kt_{CPT}=\delta_{mn}.
		\label{ortho-cpt}
		\end{equation}
	\item For $\nu=0$, where the Hamiltonian $H_0$ is Hermitian, $C=P$.
	\end{enumerate}
In Eq.~(\ref{cpt-inner}), $\gamma $ is the contour in the complex plane used to impose the vanishing boundary conditions for the eigenvalue problem of (\ref{h}), \cite{bender}. For $\nu\in[0,2)$, $\gamma $ may be taken as the real line $\R$. 

The purpose of this article is two fold. Firstly, we show that the results of Ref.~\cite{bbj}, in particular the items 1-4 of the above list, may be explained as a straightforward application of the theory of pseudo-Hermitian operators, \cite{p1,p2,p3,p4,p7}. Secondly, we outline an extension of these results to the class of quasi-Hermitian Hamiltonians, i.e., diagonalizable Hamiltonians with a real spectrum\footnote{By definition, a quasi-Hermitian operator is an operator obtained from a Hermitian operator by a similarity transformation  \cite{quasi}. Therefore it is diagonalizable and has a real spectrum. The converse of this statement is also true; as shown in \cite{p2}, a diagonalizable operator with a real discrete spectrum is related to a Hermitian operator through a similarity transformation.}, and more generally diagonalizable pseudo-Hermitian Hamiltonians. In order to achieve this purpose, we explore certain symmetry properties and involution operators associated with pseudo-Hermitian Hamiltonians.

The organization of the article is as follows. In Section~2, we offer a discussion of pseudo-Hermitian operators and their symmetries. In Section~3, we consider the problem of the existence and characterization of certain involution operators associated with a pseudo-Hermitian Hamiltonian. In Section~4, we explain the mathematical structure underlying the results of \cite{bbj} for the Hamiltonians~(\ref{h}) with $\nu\in[0,2)$ and $\gamma =\R$. In Section~5, we introduce generalized parity (${\cal P}$), time-reversal (${\cal T}$), and charge-conjugation (${\cal C}$) operators for an arbitrary quasi-Hermitian Hamiltonian $H$ with a discrete spectrum and establish the  ${\cal PT}$-  and 
${\cal CPT}$-invariance of $H$. In Section~6, we extend the results of Section~5 to the more general class of diagonalizable pseudo-Hermitian operators with a discrete spectrum. Finally in Section~7, we conclude the article with a summary of our main results.

\section{Pseudo-Hermitian Operators and Their Symmetries}

A linear operator $H$ acting in a Hilbert space ${\cal H}$ is said to be pseudo-Hermitian \cite{p1} if there is a linear, invertible, Hermitian operator $\eta:{\cal H}\to {\cal H}$ such that 
	\be
	H^\dagger=\eta H\eta^{-1}.
	\label{ph}
	\end{equation}
For a given pseudo-Hermitian operator $H$, $\eta$ is not unique \cite{p4,p8}. If one fixes a particular $\eta$ one says that $H$ is $\eta$-pseudo-Hermitian. In this case, $H$ is Hermitian with respect to the pseudo-inner product\footnote{We use the term pseudo-inner product for a possibly (but not necessarily) indefinite inner product.}
	\be
	\bbr\phi|\psi\kkt_\eta:=\br\phi|\eta\psi\kt,
	\label{eta-inner}
	\end{equation}
where $\br~|~\kt$ is the inner product of ${\cal H}$.

For diagonalizable Hamiltonians with a discrete spectrum pseudo-Hermiticity is equivalent to the condition that the complex eigenvalues come in complex-conjugate pairs \cite{p1}. Here the discreteness of the spectrum is not essential, and as shown in \cite{p7} the diagonalizability condition may be replaced by a weaker block-diagonalizability condition. Furthermore, for the class of diagonalizable Hamiltonians with a discrete spectrum, pseudo-Hermiticity is also equivalent to the condition that the Hamiltonian admits an antilinear symmetry \cite{p3}. 

Pseudo-Hermiticity also provides a characterization of the reality of the spectrum for diagonalizable Hamiltonians with a discrete spectrum. Specifically it may be used to establish the equivalence of the following statements \cite{p2}.
	\begin{enumerate}
	\item The spectrum is real;
	\item The Hamiltonian is quasi-Hermitian;
	\item Among the operators $\eta$ satisfying (\ref{ph}) there is a positive operator $\eta_+$, i.e., the Hamiltonian
	 is $\eta_+$-pseudo-Hermitian for a positive operator $\eta_+$;
	\item The Hamiltonian is Hermitian with respect to a positive-definite inner product, namely
	$\bbr~|~\kkt_{\eta_+}$, \cite{p3,p4}. 
	\end{enumerate}
One can actually construct $\eta_+$. Given a quasi-Hermitian Hamiltonian $H$ and an associated complete biorthonormal system $\{|\psi_n,a\kt,|\phi_n,a\kt\}$, which by definition satisfies 
$H|\psi_n,a\kt=E_n|\psi_n,a\kt$, $H^\dagger|\phi_n,a\kt=E^*_n|\phi_n,a\kt$, and
	\bea
	&&\br\phi_n,a|\psi_m,b\kt=\delta_{nm}\delta_{ab},
	\label{bi-ortho}\\
	&&\sum_n\sum_{a=1}^{d_n}|\psi_n,a\kt\br\phi_n,a|=1,
	\label{complete}\\
	&&H=\sum_n \sum_{a=1}^{d_n}E_n|\psi_n,a\kt\br\phi_n,a|,
	\label{sr}
	\eea
one can express $\eta_+$ according to
	\be
	\eta_+=\sum_n\sum_{a=1}^{d_n} |\phi_n,a\kt\br\phi_n,a|.
	\label{eta_+}
	\end{equation}
In Eqs.~(\ref{bi-ortho}) -- (\ref{eta_+}) and throughout this article $n$ and $m$ are spectral labels taking nonnegative integer values, $d_n$ stands for the multiplicity or degree of degeneracy of $E_n$, and $a$ and $b$ are degeneracy labels.

It turns out that $\eta_+$ is unique up to the choice of the biorthonormal system 
$\{|\psi_n,a\kt,|\phi_n,a\kt\}$, \cite{p4}. However, besides $\eta_+$, there are nonpositive invertible 
Hermitian operators $\eta$ that are also associated with the same biorthonormal system  and satisfy (\ref{ph}). These are determined by a sequence $\sigma:=\{\sigma_n^a\}$ of signs $\sigma_n^a=\pm$ and have the general form
	\be
	\eta_{\sigma}:=\sum_n \sum_{a=1}^{d_n}\sigma_n^a |\phi_n,a\kt\br\phi_n,a|.
	\label{eta=}
	\end{equation}
Obviously, the choice of the biorthonormal system is arbitrary. This means that given a complete biorthonormal system $\{|\psi_n,a\kt,|\phi_n,a\kt\}$, we can express the most general $\eta$ satisfying (\ref{ph}) according to $(\ref{eta=})$ with $|\phi_n,a\kt$ replaced with possibly different eigenvectors of $H^\dagger$ with the same eigenvalue as $|\phi_n,a\kt$. Labeling these by $|\tilde\phi_n,a\kt$ and noting that both $|\phi_n,a\kt$ and $|\tilde\phi_n,a\kt$ form bases of ${\cal H}$, we have 
$|\tilde\phi_n,a\kt=A^\dagger|\phi_n,a\kt$ for some invertible linear operator $A:{\cal H}\to{\cal H}$. Clearly, the vectors $|\tilde\phi_n,a\kt$ and $|\tilde\psi_n,a\kt:=A^{-1}|\psi_n,a\kt$ form a complete biorthonormal system. Furthermore, the operator $A$ commutes with the Hamiltonian, and 
	\be
	\eta=A^\dagger \eta_{\sigma} A.
	\label{gen-eta}
	\end{equation}
This proves the following proposition. Here we include a direct proof for completeness.
	\begin{itemize}
	\item[] {\bf Proposition~1:} For a given quasi-Hermitian Hamiltonian $H$ with a complete biorthonormal system $\{|\psi_n,a\kt,|\phi_n,a\kt\}$, the most general Hermitian invertible linear operator $\eta$ satisfying (\ref{ph}) is given by (\ref{gen-eta}) where $A$ is an invertible linear
operator commuting with the Hamiltonian (a symmetry generator) and $\sigma=\{\sigma_n^a\}$ is
a sequence of signs $\sigma_n^a=\pm$.
	\item[] {\bf Proof:} Let $\eta$ be an arbitrary Hermitian invertible linear operator satisfying 	(\ref{ph}). Then one can easily check that $X:=\eta_+^{-1}\eta$ commutes with $H$, \cite{p1}.
	Therefore, $X$ is an invertible linear operator generating a symmetry of $H$. This implies that 	$X$ and $H$ have simultaneous eigenvectors. In particular, $X$ has the form	
		\be
		X=\sum_n\sum_{a,b=1}^{d_n} x^n_{ab} |\psi_n,a\kt\br\phi_n,b|,
		\label{a1}
		\end{equation}
	where $x^n_{ab}$ are complex coefficients.	Expressing $\eta$ in terms of $\eta_+$ and $X$ 
	and using Eqs.~(\ref{a1}), (\ref{eta_+}) and (\ref{bi-ortho}), we find
		\be
		\eta=\eta_+X=\sum_n\sum_{a,b=1}^{d_n} x^n_{ab} |\phi_n,a\kt\br\phi_n,b|.
		\label{a2}
		\end{equation}
	Taking the adjoint of both sides of this equation and making use of the Hermiticity of $\eta$, we
	have $x^{n*}_{ab}=x^{n}_{ba}$. Hence the matrices $x^n$ with entries $x^{n}_{ab}$ are
	Hermitian; they may be diagonalized: 
		\be
		x^n=u^n \;x^n_{\rm diag}\;u^{n\dagger},
		\label{a3}
		\end{equation}
	where $u^n$ are $d_n\times d_n$ unitary matrices and $x^n_{\rm diag}$ are $d_n\times d_n$ 	diagonal real matrices. Next, we introduce 
		\bea
		U&:=&\sum_n\sum_{a,b=1}^{d_n} u^n_{ab}|\psi_n,a\kt\br\phi_n,b|,\nn\\
		D&:=&\sum_n\sum_{a=1}^{d_n}\sqrt{|x^n_{a}|}\;|\psi_n,a\kt\br\phi_n,a|\nn\\
		A&:=&DU=\sum_n\sum_{a,b=1}^{d_n}
			\sqrt{|x^n_{a}|} \;u^n_{ab}|\psi_n,a\kt\br\phi_n,b|,
		\label{a4}
		\eea
	where $u^n_{ab}$ and $x^n_a$ denote the entries of $u^n$ and the diagonal entries of 
	$x^n_{\rm diag}$, respectively. Note that because $u^n$ are unitary matrices $U$ is invertible. 
	In fact one can check by direct computation that 
		\[ U^{-1}:=\sum_n\sum_{a,b=1}^{d_n} u^{*n}_{ba}|\psi_n,a\kt\br\phi_n,b|\]
	satisfies $U^{-1}U=UU^{-1}=1$. Furthermore, because $X=\eta_+^{-1}\eta$, it is invertible,
	its eigenvalues $x^n_{a}$ are nonzero, and $D$ is also invertible. This in turn 
	implies that $A$ is invertible. Finally, using Eqs.~(\ref{a2}) -- (\ref{a4}), (\ref{eta=}), 
	(\ref{eta_+}), (\ref{bi-ortho}) and setting $\sigma_n^a:=x^n_{a}/|x^n_{a}|$, we can compute
		\[A^\dagger\eta_\sigma A=\sum_n\sum_{abc} 
		u^n_{ac}x^n_c u^{n*}_{bc} |\phi_n,a\kt\br\phi_n,b|=\eta.~~~\square\]
	\end{itemize}

Another interesting property of quasi-Hermitian Hamiltonians with a discrete spectrum is that they admit an exact antilinear symmetry. This follows from the observation that every diagonalizable pseudo-Hermitian Hamiltonian with a discrete spectrum is anti-pseudo-Hermitian with respect to the antilinear operator \cite{p3}
	\be
	\tau_+:=\sum_n\sum_{a=1}^{d_n} |\phi_n,a\kt\star\br\phi_n,a|,
	\label{tau}
	\end{equation}
where $\star$ is the operation of the complex conjugation of numbers. In particular, for all 
$|\phi\kt,|\psi\kt\in{\cal H}$,
	\[\star\,\br\phi|\psi\kt:=\br\phi|\psi\kt^*=\br\psi|\phi\kt.\]
Anti-pseudo-Hermiticity of $H$ with respect to $\tau_+$ means
	\be
	H^\dagger=\tau_+ H\tau_+^{-1}.
	\label{anti-ph}
	\end{equation}
Again, up to the choice of a complete biorthonormal system, (\ref{tau}) is the unique antilinear, Hermitian, invertible operator satisfying (\ref{anti-ph}). This in turn leads to the following theorem.
Again here we include an explicit proof for completeness.
	\begin{itemize}
	\item[] {\bf Proposition~2:} For a given diagonalizable pseudo-Hermitian Hamiltonian $H$ with 	a complete biorthonormal system $\{|\psi_n,a\kt,|\phi_n,a\kt\}$, the most general antilinear, 
	Hermitian, invertible operator satisfying (\ref{anti-ph}) has the form
		\be
		\tau=A^\dagger\tau_+ A,
		\label{gen-tau}
		\end{equation}
	where $A$ is an invertible linear operator commuting with $H$.
	\item[] {\bf Proof:} Let $\tau$ be an arbitrary antilinear, Hermitian, invertible operator 	satisfying (\ref{anti-ph}). Then one can easily check that $X:=\tau_+^{-1}\tau$ commutes with 	$H$, \cite{p1}. Therefore, $X$ is an invertible linear operator generating a symmetry of $H$ and
	having the form (\ref{a1}). Solving for $\tau$ in $X:=\tau_+^{-1}\tau$ and using 
	(\ref{tau}), (\ref{a1}), and (\ref{bi-ortho}), we have
		\be
		\tau=\sum_n\sum_{a,b=1}^{d_n} x^{n*}_{ab} |\phi_n,a\kt\star\br\phi_n,b|.
		\label{b1}
		\end{equation}
	Now, we recall that $\tau$ is a Hermitian antilinear operator. Therefore 
	$\br\psi_n,a|\tau|\psi_n,b\kt=\br\psi_n,b|\tau|\psi_n,a\kt$. Substituting (\ref{b1}) in this equation
	we find $x^{n}_{ab}=x^{n}_{ba}$, i.e., the matrices $x^n$ formed out of $x^n_{ab}$ are
	in general complex symmetric matrices. As shown in \cite{p6}, the latter admit a factorization of
	the form 
		\be
		x^n=a^{nT} a^{n},
		\label{b1.1}
		\end{equation}
	where $a^n$ are $n\times n$ matrices and the superscript $^T$ denotes the transpose. Next, we 
	introduce
		\be
		A:=\sum_n\sum_{a,b=1}^{d_n} a^{n}_{ab} |\psi_n,a\kt\br\phi_n,b|,
		\label{b2}
		\end{equation}
	where $a^n_{ab}$ are entries of $a^n$. Clearly, $A$ commutes with $H$. Moreover, 
	using Eqs.~(\ref{tau}), (\ref{b1}) -- (\ref{b2}), and (\ref{bi-ortho}), we have
		\[A^\dagger\tau_+ A=\sum_n\sum_{a,b=1}^{d_n} a^{n*}_{ca}a^{n^*}_{cb}
		|\phi_n,a\kt\star\br\phi_n,b|=\tau.~~~\square\]
	\end{itemize}

For a quasi-Hermitian Hamiltonian with a discrete spectrum, we can use Eq.~(\ref{gen-tau}) to define antilinear analogs of the operators $\eta_\sigma$, namely
	\be
	\tau_\sigma:=\sum_n\sum_{a=1}^{d_n}\sigma_n^a|\phi_n,a\kt\star\br\phi_n,a|,
	\label{tau=}
	\end{equation}
where again $\sigma=\{\sigma_n^a\}$ is a sequence of signs $\sigma_n^a=\pm$. This is simply done 
by setting $a^n_{ab}=(\sqrt{\sigma_n^a})^*\delta_{ab}$ in (\ref{gen-tau}).

Combining Eqs.~(\ref{ph}) and (\ref{anti-ph}), we see that $H$ commutes with
	\be
	{\cal X}:=\eta^{-1}\tau,
	\label{X=}
	\end{equation}
where $\eta$ and $\tau$ are linear and antilinear Hermitian, invertible, operators such that $H$ is 
$\eta$-pseudo-Hermitian and $\tau$-anti-pseudo-Hermitian; they have the general form (\ref{gen-eta}) and (\ref{gen-tau}), respectively. In particular if we set $\eta=\eta_{\sigma}$ and $\tau=\tau_+$ in (\ref{X=}), we find a set of {\em canonical antilinear symmetry generators}:
	\be
	{\cal X}_{\sigma}:=\eta_{\sigma}^{-1}\tau_+=\eta_+^{-1}\tau_\sigma.
	\label{can-X}
	\end{equation} 

In view Eqs.~(\ref{bi-ortho}), (\ref{tau=}), and the identity \cite{p1}:
	\be 
	\eta_{\sigma}^{-1}=\sum_n \sum_{a=1}^{d_n}\sigma_n^a |\psi_n,a\kt\br\psi_n,a|,
	\label{can-eta-inv}
	\end{equation}
we can easily calculate 
	\be 
	{\cal X}_{\sigma}=\sum_n\sum_{a=1}^{d_n}\sigma_n^a|\psi_n,a\kt\star\br\phi_n,a|.
	\label{can-X=}
	\end{equation}
It is not difficult to show that in view of (\ref{sr}), (\ref{bi-ortho}), and (\ref{can-X=}),
	\bea
	[{\cal X}_{\sigma}, H]&=&0,
	\label{sym}\\
	{\cal X}_{\sigma}|\psi_n,a\kt&=&\sigma_n^a|\psi_n,a\kt.
	\label{exact}
	\eea
In particular,
	\be
	{\cal X}_+:=\eta_+^{-1}\tau_+=\eta_{\sigma}^{-1}\tau_\sigma,
	\label{+sym}
	\end{equation}
satisfies
	\bea
	[{\cal X}_+, H]&=&0,
	\label{sym+}\\
	{\cal X}_+|\psi_n,a\kt&=&|\psi_n,a\kt.
	\label{exact+}
	\eea
Hence the antilinear symmetry generated by ${\cal X}_{\sigma}$ is an exact symmetry. The converse of this statement is also valid. That is if a diagonalizable Hamiltonian with a discrete spectrum admits an exact symmetry generated by an invertible antilinear operator, then its spectrum is real \cite{p2}; it is quasi-Hermitian. A direct consequence of this statement is that if a diagonalizable pseudo-Hermitian Hamiltonian with a discrete spectrum has nonreal eigenvalues, then it cannot support exact antilinear symmetries. Such a Hamiltonian always admits antilinear symmetries \cite{p3}, but these symmetries are necessarily broken. 

We can repeat the above analysis of quasi-Hermitian Hamiltonians for the more general diagonalizable pseudo-Hermitian Hamiltonians with a discrete spectrum \cite{p1,p3}. For the latter Hamiltonians nonreal eigenvalues come in complex-conjugate pairs with identical multiplicity, so we identify the 
spectral label $n$ with $\nu_0$, $\nu_+$, or $\nu_-$ depending on whether the imaginary part of 
$E_n$ is zero, positive, or negative, respectively. In this case, Eqs.~(\ref{bi-ortho}) -- (\ref{sr}), with $n=\nu_0,\nu_\pm$ and $m=\mu_0,\mu_\pm$, are still valid, $d_{\nu_+}=d_{\nu_-}$, and the 
analog of the positive operator (\ref{eta_+}) is the operator
	\be
	\eta_+=\sum_{\nu_0}\sum_{a=1}^{d_{\nu_0}}|\phi_{\nu_0},a\kt\br\phi_{\nu_0},a|+
		\sum_{\nu}\sum_{a=1}^{d_{\nu}}(|\phi_{\nu_+},a\kt\br\phi_{\nu_-},a|+
		|\phi_{\nu_-},a\kt\br\phi_{\nu_+},a|).
	\label{eta_+2}
	\end{equation}
Here we use $\nu$ to denote the common value of $\nu_\pm$.

It is not difficult to see that the proof of Proposition~1 extends to the class of diagonalizable pseudo-Hermitian Hamiltonians with a discrete spectrum; it yields the following generalization of 
Proposition~1, see also \cite{p3}.
	\begin{itemize}
	\item[] {\bf Proposition~3:} For a given diagonalizable pseudo-Hermitian Hamiltonian $H$ with
a complete biorthonormal system $\{|\psi_n,a\kt,|\phi_n,a\kt\}$, the most general Hermitian invertible linear operator $\eta$ satisfying (\ref{ph}) is given by (\ref{gen-eta}) where $A$ is an invertible linear operator commuting with the Hamiltonian,
	\be
	\eta_{\sigma}:=\sum_{\nu_0}\sum_{a=1}^{d_{\nu_0}}\sigma_{\nu_0}^a
		|\phi_{\nu_0},a\kt\br\phi_{\nu_0},a|+
		\sum_{\nu}\sum_{a=1}^{d_{\nu}}(|\phi_{\nu_+},a\kt\br\phi_{\nu_-},a|+
		|\phi_{\nu_-},a\kt\br\phi_{\nu_+},a|),
	\label{eta_c}
	\end{equation}
and $\sigma=\{\sigma_{\nu_0}^a\}$ is a sequence of signs $\sigma_{\nu_0}^a=\pm$. 	
	\end{itemize}
Similarly, one can show that every diagonalizable pseudo-Hermitian Hamiltonian $H$ with a discrete spectrum admits antilinear symmetries generated by (\ref{X=}). For instance we have the canonical antilinear symmetry generators (\ref{can-X}) where now $\eta_+$ is given by (\ref{eta_+2}) and
	\be
	\tau_\sigma:=\sum_{\nu_0}\sum_{a=1}^{d_{\nu_0}}
		\sigma_{\nu_0}^a	|\phi_{\nu_0},a\kt\star\br\phi_{\nu_0},a|+
		\sum_{\nu}\sum_{a=1}^{d_{\nu}}
		(|\phi_{\nu_+},a\kt\star\br\phi_{\nu_+},a|+
		|\phi_{\nu_-},a\kt\star\br\phi_{\nu_-},a|).
	\label{tau_+2}
	\end{equation}
We can express these symmetry generators according to
	\be 
	{\cal X}_{\sigma}=\sum_{\nu_0} \sum_{a=1}^{d_{\nu_0}}
	\sigma_{\nu_0}^a |\psi_{\nu_0},a\kt\star\br\phi_{\nu_0},a|+
	\sum_{\nu} \sum_{a=1}^{d_{\nu}} (|\psi_{\nu_+},a\kt\star\br\phi_{\nu_-},a|+
	|\psi_{\nu_-},a\kt\star\br\phi_{\nu_+},a|),
	\label{gen-can-X=}
	\end{equation}
where we have used Eqs.~(\ref{can-X}), (\ref{eta_+2}), (\ref{tau_+2}), (\ref{bi-ortho})
and the identity \cite{p1}
	\be 
	\eta_{\sigma}^{-1}=\sum_{\nu_0}\sum_{a=1}^{d_{\nu_0}}\sigma_{\nu_0}^a
		|\psi_{\nu_0},a\kt\br\psi_{\nu_0},a|+
		\sum_{\nu}\sum_{a=1}^{d_{\nu}}(|\psi_{\nu_+},a\kt\br\psi_{\nu_-},a|+
		|\psi_{\nu_-},a\kt\br\psi_{\nu_+},a|).
	\label{gen-can-eta-inv}
	\end{equation}

Next, we observe that in light of Eqs.~(\ref{sr}), (\ref{bi-ortho}), and (\ref{gen-can-X=}),
	\bea
	[{\cal X}_{\sigma}, H]&=&0,
	\label{gen-sym}\\
	{\cal X}_{\sigma}|\psi_n,a\kt&=&\left\{\begin{array}{ccc}
	\sigma_{\nu_0}^a|\psi_{\nu_0},a\kt&{\rm  if}&n=\nu_0\\
	|\psi_{\nu_\mp},a\kt&{\rm  if}&n=\nu_\pm.\end{array}\right.
	\label{broken}
	\eea
In particular, the operator~(\ref{+sym}) satisfies
	\bea
	[{\cal X}_+, H]&=&0,
	\label{gen-sym+}\\
	{\cal X}_+|\psi_n,a\kt&=&\left\{\begin{array}{ccc}
	|\psi_{\nu_0},a\kt&{\rm  if}&n=\nu_0\\
	|\psi_{\nu_\mp},a\kt&{\rm  if}&n=\nu_\pm.\end{array}\right.
	\label{broken+}
	\eea
Therefore, ${\cal X}_{\sigma}$ generate symmetries of $H$ which are however broken.

\section{Involution Operators Associated with a Pseudo-\\ Hermitian Hamiltonian}

Among the basic properties of the $P$, $T$, and $PT$ operators (within the scalar/bosonic quantum mechanics) is that they are involutions of the Hilbert space, i.e., their square is the identity operator. In this section we study the problem of the existence and characterization of certain involutions of the Hilbert space which are associated with a given pseudo-Hermitian Hamiltonian. 
	\begin{itemize}
	\item[] {\bf Proposition~4:} The operators $S_\sigma:=\eta_+^{-1}\eta_\sigma$ and
${\cal X}_\sigma:=\eta_+^{-1}\tau_\sigma$ are involutions.
	\item[] {\bf Proof:} according to Eqs.~(\ref{eta_+}), (\ref{eta=}), and 
(\ref{bi-ortho}), we have
	\be
	S_\sigma=\sum_{\nu_0}\sum_{a=1}^{d_{\nu_0}}\sigma_{\nu_0}^a
		|\psi_{\nu_0},a\kt\br\phi_{\nu_0},a|+
		\sum_{\nu}\sum_{a=1}^{d_{\nu}}(|\psi_{\nu_+},a\kt\br\phi_{\nu_-},a|+
		|\psi_{\nu_-},a\kt\br\phi_{\nu_+},a|).
	\label{S=}
	\end{equation}
Squaring this expression and using (\ref{bi-ortho}), we find $S_\sigma^2=1$. Similarly, we have
${\cal X}_\sigma^2=1$.~~~$\square$
	\item[] {\bf Corollary~1:} Every diagonalizable pseudo-Hermitian Hamiltonian $H$ with a discrete spectrum admits a symmetry generated by a linear involution $S$ and a symmetry generated by an antilinear involution $\Sigma$, i.e., $[H,S]=[H,\Sigma]=0$  and $S^2=\Sigma^2=1$.
	\item[] {\bf Proof:} Again we recall that because $\eta_+$ and $\eta_\sigma$ satisfying (\ref{ph}), the linear operator $S=S_\sigma:=\eta_+^{-1}\eta_\sigma$ commutes with the Hamiltonian \cite{p1}. Therefore, in view of Proposition~4, $S$ and $\Sigma={\cal X}_\sigma$ are involutions generating symmetries of $H$. Clearly, $S$ is linear whereas $\Sigma$ is antilinear.~~~$\square$
	\item[] {\bf Corollary~2:} Let $H$ be a diagonalizable Hamiltonian with a discrete spectrum. Then $H$ is pseudo-Hermitian if and only if it admits an antilinear symmetry generated by an involution.
	\item[] {\bf Proof:} If $H$ is pseudo-Hermitian, then according to Proposition~4 it admits such a symmetry. Conversely, suppose that $H$ admits such a symmetry. Then because this is an antilinear symmetry, $H$ must be pseudo-Hermitian, \cite{p3}.~~~$\square$
	\item[] {\bf Proposition~5:}  A diagonalizable Hamiltonian $H$ with a discrete spectrum is 
	anti-pseudo-Hermitian with respect to a Hermitian antilinear involution if and only if there 
	is a complete biorthonormal system $\{|\psi_n,a\kt,|\phi_n,a\kt\}$ satisfying
		\be
		\br\phi_n,a|\phi_m,b\kt=\br\psi_m,b|\psi_n,a\kt.
		\label{inv-tau-condi}
		\end{equation}
	\item[] {\bf Proof:} Suppose $H$ is anti-pseudo-Hermitian with respect to a Hermitian 	antilinear involution $\tau$. Then there is a complete biorthonormal system 
	$\{|\psi_n,a\kt,|\phi_n,a\kt\}$ for which $\tau=\tau_+$. Now, imposing the condition that 
	$\tau^2=1$ and using Eq.~(\ref{bi-ortho}), one finds (\ref{inv-tau-condi}). Conversely, one can 	check that if a complete biorthonormal system $\{|\psi_n,a\kt,|\phi_n,a\kt\}$ satisfies this 
	equation, the Hermitian antilinear operator $\tau_+$ given by (\ref{tau}) is an involution. As we 	mentioned above and shown in \cite{p3}, $H$ is anti-pseudo-Hermitian with respect to this
	operator.~~~$\square$
	\item[] {\bf Corollary~3:} A pseudo-Hermitian Hamiltonian $H$ is anti-pseudo-Hermitian with 	respect to a Hermitian antilinear involution if and only if for every complete biorthonormal 	system $\{|\psi_n,a\kt,|\phi_n,a\kt\}$ there is an invertible linear symmetry generator $A$
	satisfying
		\be
		\sum_n\sum_{a=1}^{d_n}\br\psi_k,c|(AA^\dagger)^{-1}|\psi_n,a\kt
		\br\psi_m,b|(AA^\dagger)^{-1}|\psi_n,a\kt=\delta_{km}\delta_{bc}.
		\label{inv-tau-condi-2}
		\end{equation}
	\item[] {\bf Proof:} According to Proposition~5 anti-pseudo-Hermiticity of $H$ with respect to
	a Hermitian antilinear involution is equivalent to the existence of a complete 
	biorthonormal system $\{|\tilde\psi_n,a\kt,|\tilde\phi_n,a\kt\}$ satisfying 
		\be
		\br\tilde\phi_n,a|\tilde\phi_m,b\kt=\br\tilde\psi_m,b|\tilde\psi_n,a\kt.
		\label{c1}
		\end{equation}
	Now, let $\{|\psi_n,a\kt,|\phi_n,a\kt\}$ be an arbitrary complete biorthonormal system. Then 
	there is a linear invertible symmetry generator $A$ satisfying 
	$|\tilde\psi_n,a\kt=A^{-1}|\psi_n,a\kt$ and $|\tilde\phi_n,a\kt=A^\dagger|\phi_n,a\kt$. 
	Substituting these relations in (\ref{c1}), we find
		\be
		\br\phi_n,a|AA^\dagger|\phi_m,b\kt=\br\psi_m,b|(AA^\dagger)^{-1}|\psi_n,a\kt.
		\label{c1.1}
		\end{equation}
	Next, we multiply $\br\psi_k,c|(AA^\dagger)^{-1}|\psi_n,a\kt$ by both sides of (\ref{c1.1})
	and sum over $n$ and $a$. This yields (\ref{inv-tau-condi-2}). Conversely, 
	assuming the existence of an invertible symmetry generator $A$ satisfying 
	(\ref{inv-tau-condi-2}), one can easily check that the complete 
	biorthonormal system defined by $|\tilde\psi_n,a\kt:=A^{-1}|\psi_n,a\kt$ and 
	$|\tilde\phi_n,a\kt:=A^\dagger|\psi_n,a\kt$ satisfies~(\ref{c1}).~~~$\square$
	\end{itemize}
Eq.~(\ref{c1.1}) is particularly useful as it gives the necessary and sufficient conditions for a given 
invertible Hermitian antilinear operator $\tau$ satisfying (\ref{anti-ph}) to be an involution. For example, in order to find the necessary and sufficient conditions under which $\tau_\sigma$ of 
Eq.~(\ref{tau_+2}) is an involution, we write $\tau_\sigma=A^\dagger\tau_+ A$, where
	\[ A=\sum_{\nu_0}\sum_{a=1}^{d_{\nu_0}}(\sqrt{\sigma_{\nu_0}^a})^*\;|\psi_{\nu_0},a\kt
	\br\phi_{\nu_0},a|+\sum_\nu\sum_{a=1}^{d_\nu}(
	|\psi_{\nu_+},a\kt\br\phi_{\nu_+},a|+|\psi_{\nu_-},a\kt\br\phi_{\nu_-},a|),\]
and substitute this equation in (\ref{c1.1}). This yields the following conditions
	\bea
	\br\phi_{\nu_0},a|\phi_{\mu_0},b\kt&=&\sigma_{\nu_0}^a\sigma_{\mu_0}^b
		\br\psi_{\mu_0},b|\psi_{\nu_0},a\kt,
	\label{c3.1}\\
	\br\phi_{\nu_0},a|\phi_{\mu_\pm},b\kt&=&\sigma_{\nu_0}^a
		\br\psi_{\mu_\pm},b|\psi_{\nu_0},a\kt,
	\label{c3.2}\\
	\br\phi_{\nu_\pm},a|\phi_{\mu_\pm},b\kt&=&
		\br\psi_{\mu_\pm},b|\psi_{\nu_\pm},a\kt.
	\label{c3.3}
	\eea

	\begin{itemize}
	\item[] {\bf Proposition~6:}  A diagonalizable Hamiltonian $H$ with a discrete spectrum is 
	pseudo-Hermitian with respect to a Hermitian linear involution $\eta$ if and only if there is a 	complete biorthonormal system $\{|\psi_n,a\kt,|\phi_n,a\kt\}$, with $n=\nu_0,\nu_\pm$ as 
	above, and a sequence of signs $\sigma=\{\sigma_{\nu_0}^a\}$ such that
		\bea
		\br\phi_{\nu_0},a|\phi_{\mu_0},b\kt&=&
		\sigma_{\nu_0}^a\sigma_{\mu_0}^b\br\psi_{\nu_0},a|\psi_{\mu_0},b\kt,
		\label{inv-eta-condi-1}\\
		\br\phi_{\nu_0},a|\phi_{\mu_\pm},b\kt&=&
		\sigma_{\nu_0}^a \br\psi_{\nu_0},a|\psi_{\mu_\mp},b\kt,
		\label{inv-eta-condi-2}\\
		\br\phi_{\nu_\pm},a|\phi_{\mu_\pm},b\kt&=&\br\psi_{\nu_\mp},a|\psi_{\mu_\mp},b\kt.
		\label{inv-eta-condi-3}
		\eea
	\item[] {\bf Proof:} This follows from a similar argument as the one used in the proof of 	Proposition~5. It is based on the observation that $\eta$ takes the canonical form~(\ref{eta_c})
	in some complete biorthonormal system $\{|\psi_n,a\kt,|\phi_n,a\kt\}$ and that in this system
	the condition $\eta^2=1$ is equivalent to Eqs.~(\ref{inv-eta-condi-1}) -- 
	(\ref{inv-eta-condi-3}).~~~$\square$
	\item[] {\bf Corollary~4:} Let $H$ be a diagonalizable  pseudo-Hermitian Hamiltonian $H$ 
	with a discrete spectrum and a complete biorthonormal system $\{|\psi_n,a\kt,|\phi_n,a\kt\}$.
	Then the operators $\tau_\sigma$ of (\ref{tau_+2}) and $\eta_\sigma$ of (\ref{eta_c}) are
	involutions if and only if Eqs.~(\ref{c3.1}) -- (\ref{c3.3}) and 
	(\ref{inv-eta-condi-1}) -- (\ref{inv-eta-condi-3}) are satisfied. Furthermore, in this case 
		\be
		[\tau_\sigma,\eta_\sigma]=0.
		\label{commu}
		\end{equation}
	\item[] {\bf Proof:} The equivalence of Eqs.~(\ref{c3.1}) -- (\ref{c3.3}) and 
	(\ref{inv-eta-condi-1}) -- (\ref{inv-eta-condi-3}) with the condition that $\tau_\sigma$ and
	$\eta_\sigma$ are involutions follows from Corollary~4 and Proposition~6. Finally, in view of
	the identities: $\tau_\sigma=\tau_\sigma^{-1}$, $\eta_\sigma=\eta_\sigma^{-1}$,
		\be
		\tau_+^{-1}=\sum_n\sum_{a=1}^{d_n} |\psi_n,a\kt\star\br\psi_n,a|,
		\label{tau-inv}
		\end{equation}
	and Eqs.~(\ref{tau_+2}), (\ref{eta_c}), (\ref{gen-can-eta-inv}), (\ref{bi-ortho}),
	we have
		\[\tau_\sigma\eta_\sigma=\tau_\sigma\eta_\sigma^{-1}=\tau_+\eta_+^{-1}
		=\eta_+\tau_+^{-1}=\eta_\sigma\tau_\sigma.~~~\square\]
	\end{itemize}

\section{Application to Hamiltonians~(\ref{h}) with $\gamma =\R$}

Consider the class of $PT$-symmetric Hamiltonians $H_\nu$ of Eq.~(\ref{h}) with $\nu\in[0,2)$, $\gamma =\R$, and ${\cal H}=L^2(\R)$. Then, following \cite{bbj}, we may choose a set of eigenvectors\footnote{Note that what we denote by $|\phi_n\kt$ are eigenvectors of $H^\dagger$. This is the notation used in \cite{p1,p2,p3,p4,p7} which differs from that of \cite{bbj}.} $|\psi_n\kt$ of 
$H_\nu$ satisfying
	\be
	PT|\psi_n\kt =|\psi_n\kt.
	\label{pt=}
	\end{equation}
Because the eigenvalues of $H_\nu$ are nondegenerate, we have dropped the degeneracy label $a=1$.
Also as usual the $PT$ operator is defined by $PT\psi(x):=[\psi(-x)]^*$ where $|\psi\kt$ is an arbitrary state vector represented by the wave function $\psi(x)$. Moreover, relying on the numerical evidence \cite{bender} that is also used in \cite{bbj}, we assume the validity of the completeness relation 
	\be
	\sum_n (-1)^n\psi_n(x)\psi_n(y)=\delta(x-y),
	\label{comp1}
	\end{equation}
and the orthogonality condition
	\be
	(\psi_m,\psi_n)=(-1)^n\delta_{mn},
	\label{ortho-1}
	\end{equation}
where the indefinite inner product $(~,~)$ is defined by
	\be
	(\phi,\psi):=\int_{\R} dx\; [PT \phi(x)]\psi(x).
	\label{pt-inner}
	\end{equation}

Introducing the functions
	\be
	\phi_n(x):=(-1)^n\psi_n(x)^*,
	\label{phi=}
	\end{equation}
which also belong to ${\cal H}=L^2(\R)$, and using Eqs.~(\ref{ortho-1}) and (\ref{pt-inner})
we can show that
	\[\br\phi_m|\psi_n\kt:=\int_{\R}dx\;\phi_m(x)^*\psi_n(x)=
	(-1)^m(\psi_m,\psi_n)=\delta_{mn}.\]
This coincides with the biorthonormality relation~(\ref{bi-ortho}). Furthermore, we write 
Eq.~(\ref{comp1}) in the form 
	\[\delta(x-y)=\sum_n \phi_n(x)\psi_n(y)^*=\sum_n \br x|\phi_n\kt\br\psi_n|y\kt\]
which is equivalent to the completeness relation~(\ref{complete}). 
Therefore, $\{|\psi_n\kt,|\phi_n\kt\}$ forms a complete biorthonormal system, and the $PT$-symmetric
Hamiltonians~(\ref{h}) are diagonalizable, \cite{p3}. Moreover, because their spectrum is real and discrete, these Hamiltonians are examples of quasi-Hermitian Hamiltonians having a discrete spectrum.

Next, we calculate
	\be
	(\phi,\psi)=\int_{\R} dx\; \phi(-x)^*\psi(x)=\int_{\R} dx\; \phi(x)^*\psi(-x)=
	\int_{\R} dx\; \phi(x)^*P\psi(x)=\br\phi|P|\psi\kt=\bbr\phi|\psi\kkt_P,
	\label{eq0}
	\end{equation}
where
	\be
	\br\phi|\psi\kt:=\int_{\R}dx\;\phi(x)^*\psi(x).
	\label{L2}
	\end{equation}
According to Eq.~(\ref{eq0}), the inner product (\ref{pt-inner}) is nothing but $\bbr~|~\kkt_P$. This observation together with Eqs.~(\ref{ortho-1}) and (\ref{phi=}) imply
	\be
	P=\sum_n (-1)^n|\phi_n\kt\br\phi_n|.
	\label{P-ph}
	\end{equation}
Comparing this equation with (\ref{eta=}), we see that $P$ is an example of the canonical operators
$\eta_{\sigma}$ of Eq.~(\ref{eta=}) with 
	\be
	\sigma_n=(-1)^n.
	\label{sigma=}
	\end{equation}
This is another verification of the fact that the Hamiltonians (\ref{h}) are $P$-pseudo-Hermitian, 
\cite{p1}.

Note that as a result of Eq.~(\ref{pt=}), $\psi_n(-x)^*=\psi_n(x)$. This equation together with 
(\ref{phi=}) imply
	\bea
	\phi_n(x)&=&(-1)^n\psi_n(-x),
	\label{eq01}\\
	\br\psi_m|\psi_n\kt&=&\int_{\R}dx\;\psi_m(x)^*\psi_n(x)=
	\int_{\R}dx\;\psi_m(-x)\psi_n(-x)^*\nn\\
	&=&\int_{\R}dx\;\psi_m(x)\psi_n(x)^*=\br\psi_n|\psi_m\kt,
	\label{eq02}\\
	\br\phi_m|\phi_n\kt&=&\int_{\R}dx\;\phi_m(x)^*\phi_n=(-1)^{m+n}
	\int_{\R}dx\;\psi_m(-x)^*\psi_n(-x)\nn\\
	&=&(-1)^{m+n}\int_{\R}dx\;\psi_m(x)^*\psi_n(x)
	=(-1)^{m+n}\br\psi_m|\psi_n\kt.
	\label{eq03}
	\eea
In view of Eqs.~(\ref{sigma=}) and (\ref{eq03}), the condition (\ref{inv-eta-condi-1}) of Propositions~6 holds. Therefore, Eq.~(\ref{P-ph}) is consistent with the fact that $P$ is an involution.

Next, we use Eqs.~(\ref{pt=}) and (\ref{bi-ortho}) to calculate
	\be
	PT=\sum_n |\psi_n\kt\, \star\, \br\phi_n|.
	\label{PT=}
	\end{equation}
Then multiplying both sides of this equation by $P$ and using Eqs.~(\ref{P-ph}) and (\ref{bi-ortho}), we find
	\be
	T=\sum_n (-1)^n |\phi_n\kt\, \star\, \br\phi_n|.
	\label{T=}
	\end{equation}
This shows that the time-reversal operator $T$ is nothing but the canonical antilinear operator 
(\ref{tau=}) with $\sigma_n$ given by (\ref{sigma=}).\footnote{This is consistent with the known
fact \cite{p3} that the $PT$-symmetric standard Hamiltonians of the form $H=p^2+V(x;t)$ which
have $\R$ as their configuration space, in general, and the Hamiltonians (\ref{h}) with $\nu\in[0,2)$
and $\gamma =\R$, in particular, are $T$-anti-pseudo-Hermitian. See also \cite{p5}.} Again, in view
of (\ref{eq02}) and (\ref{eq03}), we see that the condition~(\ref{c3.1}) of Corollary~4 is satisfied and 
the expression~(\ref{T=}) is consistent with $T^2=1$.

Next, we consider the positive operator $\eta_+$ for the Hamiltonians~(\ref{h}) with $\nu\in[0,2)$ and $\gamma =\R$. Because these Hamiltonians are pseudo-Hermitian with respect to both $\eta_+$ and $P$, they admit a symmetry generated by $\eta_+^{-1}P$. This is a particular example of the symmetry generators $S$ of Proposition~4, where $\nu_0=n$, $\sigma_{\nu_0}=(-1)^{\nu_0}=(-1)^n$, and $\nu_\pm$ are absent. We can compute $\eta_+^{-1}P$ using Eq.~(\ref{S=}). Alternatively, we may use the identity \cite{p1}
	\be
	\eta_+^{-1}=\sum_n |\psi_n\kt \br\psi_n|.
	\label{eta-inv}
	\end{equation}
together with Eqs.~(\ref{P-ph}) and (\ref{bi-ortho}). This yields
	\be
	\eta_+^{-1}P=\sum_n (-1)^n |\psi_n\kt\br\phi_n|.
	\label{C}
	\end{equation}
The symmetry generator $\eta_+^{-1}P$ has the following form in the position representation.
	\be
	\br x|\eta_+^{-1}P|y\kt=\sum_n (-1)^n \psi_n(x)^*\phi_n(y)=\sum_n\psi_n(x)\psi_n(y).
	\label{C2}
	\end{equation}
Comparing this equation with Eq.~(\ref{x-rep}), we see that $\eta_+^{-1}P$ coincides with the charge-conjugation operator $C$ of Ref.~\cite{bbj},
	\be
	C=\eta_+^{-1}P.
	\label{C=def}
	\end{equation}

Next, we use Eqs.~(\ref{eta-inv}), (\ref{T=}), (\ref{bi-ortho}), (\ref{eq03}), (\ref{eq02}), and 
(\ref{complete}) to compute 
	\bea
	T\eta_+&=&\sum_{nm} (-1)^n|\phi_n\kt\,\star\,\br\phi_n|\phi_m\kt\br\phi_m|
	=\sum_{nm} (-1)^m|\phi_n\kt\,\star\,\br\psi_n|\psi_m\kt\br\phi_m|\nn\\
	&=& \sum_{nm} (-1)^m|\phi_n\kt\,\star\,\br\psi_m|\psi_n\kt\br\phi_m|
	= \sum_{nm} (-1)^m|\phi_n\kt\br\psi_n|\psi_m\kt\,\star\,\br\phi_m|\nn\\
	&=& \sum_{m} (-1)^m|\psi_m\kt\,\star\,\br\phi_m|
	= \sum_{nm} (-1)^m|\psi_m\kt\br\psi_m|\phi_n\kt\,\star\,\br\phi_n|\nn\\
	&=&\eta_+^{-1}T=\eta_+^{-1}P^2T=CPT.
	\label{eq05}
	\eea
Hence,
	\bea
	\br\phi|\psi\kt_{CPT}&=&\int_{\R} dx\, [CPT \phi(x)]\psi(x)=
		\int_{\R} dx\, [T\eta_+\phi(x)]\psi(x)\nn\\
		&=&\int_{\R} dx\, [\eta_+\phi(x)]^*\psi(x)=
		\int_{\R} dx\, \phi(x)^*[\eta_+\psi(x)]\nn\\
		&=&\br\phi|\eta_+\psi\kt=\bbr\phi|\psi\kkt_{\eta_+},
	\label{eq06}
	\eea
where we have used the fact that $\eta_+$ is Hermitian. Eqs.~(\ref{eq06}) show that the $CPT$-inner product (\ref{cpt-inner}) advocated in \cite{bbj} is nothing but the positive-definite inner product $\bbr~|~\kkt_{\eta_+}$ that was extensively used in \cite{p8}. Moreover, the orthonormality relation~(\ref{ortho-cpt}) is a simple consequence of 
Eqs.~(\ref{eta=}) and (\ref{bi-ortho}).

Comparing the expressions given in (\ref{PT=}) and (\ref{eq05}) for the $PT$ and $CPT$ operators 
with Eq.~(\ref{can-X=}), we see that the $PT$ and $CPT$ operators are specific examples of the canonical antilinear symmetry generators (\ref{can-X=}).

\section{Generalized $P$, $T$, and $C$ Operators for Quasi-Hermitian Operators}

In the preceding section we explored the mathematical basis of the charge conjugation 
operator~(\ref{x-rep}) for the Hamiltonians (\ref{h}) with the choice $\gamma =\R$ which is allowed for $\nu\in[0,2)$. In this section we will demonstrate that indeed the approach based on the theory of pseudo-Hermitian operators applies to quasi-Hermitian Hamiltonians with a discrete spectrum in general and the $PT$-symmetric Hamiltonians (\ref{h}) with  $\nu\in[0,\infty)$ in particular.

As we discussed in Section~3, every quasi-Hermitian Hamiltonian $H$ with a discrete spectrum is
$\eta_+$-pseudo-Hermitian for a positive operator $\eta_+$, and that $H$ is Hermitian with respect to the inner product $\bbr~~|~~\kkt_{\eta_+}$. This in turn implies the existence of a complete set of eigenvectors $|\psi_n,a\kt$ of $H$ such that $|\psi_n\kt$ are orthonormal with respect to 
$\bbr~~|~~\kkt_{\eta_+}$.

\begin{itemize}
	\item[] {\bf Lemma~1:}
	Let $H$, $\eta_+$, and $|\psi_n,a\kt$ be as in the preceding paragraph, and
	\bea
	|\phi_n,a\kt&:=&\eta_+|\psi_n,a\kt,
	\label{int-phi}\\
	{\cal P}&:=&\sum_n \sum_{a=1}^{d_n} (-1)^n |\phi_n,a\kt\br\phi_n,a|,
	\label{gen-P}\\
	{\cal T}&:=&\sum_n  \sum_{a=1}^{d_n} (-1)^n |\phi_n,a\kt\star\br\phi_n,a|,
	\label{gen-T}\\
	{\cal C}&:=&\sum_n  \sum_{a=1}^{d_n}(-1)^n |\psi_n,a\kt\br\phi_n,a|.
	\label{gen-C}
	\eea
Then
	\begin{enumerate}
	\item $\{|\psi_n,a\kt,|\phi_n,a\kt\}$ forms a complete biorthonormal system;
	\item $\eta_+$ satisfies (\ref{eta_+}) and 
		\be
		\eta_+^{-1}={\cal T}\eta_+{\cal T};
		\label{e=TeT}
		\end{equation}
	\item $H$ is ${\cal P}$-pseudo-Hermitian and ${\cal T}$-anti-pseudo-Hermitian;
	\item ${\cal P}{\cal T}$ and ${\cal CPT}$, which have the form
		\bea
		{\cal PT}&=&\sum_n  \sum_{a=1}^{d_n}|\psi_n,a\kt\star\br\phi_n,a|,
		\label{gen-PT}\\
		{\cal CPT}&=&\sum_n \sum_{a=1}^{d_n} (-1)^n |\psi_n,a\kt\star\br\phi_n,a|,
		\label{gen-CPT}
		\eea
	are antilinear symmetry generators and $C$ is a linear symmetry generator for $H$; the 	corresponding symmetries are exact, in particular $|\psi_n,a\kt$ satisfy
		\bea
		{\cal PT}|\psi_n,a\kt&=&|\psi_n,a\kt,
		\label{gen-PT-psi}\\
		{\cal CPT}|\psi_n,a\kt&=&{\cal C}|\psi_n,a\kt=(-1)^n|\psi_n,a\kt;
		\label{gen-CPT-psi}
		\eea
	\item ${\cal P}$, ${\cal T}$, and ${\cal C}$ satisfy	
		\bea
		&&({\cal PT})^2={\cal C}^2=1, 
		\label{nilp}\\
		&& {\cal C}=\eta_+^{-1}P={\cal T}\eta_+{\cal T}{\cal P};
		\label{C==}
		\eea
	\item The operators ${\cal P}$ and ${\cal T}$ are involutions if and only if 
		\be
		(-1)^{m+n}\br\phi_n,a|\phi_m,b\kt=\br\psi_n,a|\psi_m,b\kt=\br\psi_m,b|\psi_n,a\kt;
		\label{inv-condi}
		\end{equation}
	\item If $H$ is a Hermitian Hamiltonian, ${\cal C}^{-1}{\cal P}$ is a Hermitian invertible 
	linear operator commuting with $H$. In particular, if for all $n$ and $a$, 
	$|\phi_n,a\kt=|\psi_n,a\kt$, then ${\cal C}={\cal P}$.
	\end{enumerate}
\item[] {\bf Proof:} Statement 1 may be established by checking Eqs.~(\ref{bi-ortho}) and
(\ref{complete}) directly. Statements 2-4 follow from these equations and (\ref{gen-P}) -- 
(\ref{gen-C}). ${\cal PT}$ and ${\cal CPT}$ are respectively examples of the antilinear symmetry 
generators ${\cal X}_+$ and ${\cal X}_\sigma$. Statement~5 is a result of Proposition~4; Eq.~(\ref{C==})
may be checked by direct computation. Statement~6 is a consequence of Corollary~4. In order to prove statement~7, we introduce 
	\be
	\Lambda:=\sum_n\sum_{a=1}^{d_n}|\psi_n,a\kt\br\psi_n,a|,
	\label{Lambda}
	\end{equation}
which is clearly a Hermitian invertible linear operator commuting with $H$. Now it suffices to use 
(\ref{bi-ortho}) to establish $\Lambda{\cal P}={\cal C}$. Finally for the case that
 $|\phi_n,a\kt=|\psi_n,a\kt$, Eq.~(\ref{complete}) implies $\Lambda=1$.~~~$\square$
\end{itemize}

In view of the analogy with the systems studied in Section~3, we shall respectively call the operators ${\cal P}$, ${\cal T}$, and ${\cal C}$ the {\em generalized parity, time-reversal,} and {\em charge conjugation} operators. The following theorem follows as a direct consequence of Lemma~1.
	\begin{itemize}	
	\item[] {\bf Theorem~1:} Every diagonalizable Hamiltonian with a real discrete spectrum is invariant  under the action of the generalized charge-conjugation operator ${\cal C}$ and the combined action of the generalized parity and time-reversal symmetry (${\cal PT}$). In particular, every such Hamiltonian has exact ${\cal PT}$- and ${\cal CPT}$-symmetry.
	\end{itemize}

Clearly for the Hamiltonians (\ref{h}) with $\nu\in[0,2)$, the operators ${\cal P}$, ${\cal T}$, and 
${\cal C}$ coincide with $P, T$, and $C$. For $\nu\in[2,\infty)$, we define
the vectors $|\phi_n\kt$ according to (\ref{phi=}) so that in the position representation
	\be
	\eta_+(x,y)=\sum_n \phi_n(x)\phi_n(y)^*=\sum_n\psi_n(x)^*\psi_n(y),~~~~~~~~\forall
	x,y\in\R.
	\label{zz1}
	\end{equation}
Next, we note that Eqs.~(\ref{pt=}), (\ref{phi=}), and consequently (\ref{eq01}) also hold for 
$\nu\in[2,\infty)$. Using (\ref{eq01}) and (\ref{comp1}), we can show that in the position 
representation 
	\bea
	{\cal P}(x,y)&=&\sum_n (-1)^n\phi_n(x)\phi_n(y)^*=\sum_n (-1)^n\psi_n(-x)\psi_n(-y)^*\nn\\
	&=&\sum_n (-1)^n\psi_n(-x)\psi_n(y)=\delta(x+y)=P(x,y)~~~~~~~~\forall
	x,y\in\R,
	\label{zz2}
	\eea
i.e.,  $P$ and ${\cal P}$ have the same position representations. Furthermore, we can easily see that in view of (\ref{gen-PT}) and (\ref{pt=}), ${\cal PT}=PT$, so that $T$ and ${\cal T}$ also have the same position representations. Finally, we can employ (\ref{gen-C}) and (\ref{x-rep}) to infer that $C$ and ${\cal C}$ have the same position representations as well.

\section{Generalized $P$, $T$, and $C$ Operators for Pseudo-Hermitian Hamiltonians}

The construction of the operators ${\cal P}, {\cal T}$, and ${\cal C}$ may be easily generalized to the class of all diagonalized pseudo-Hermitian operators with a discrete spectrum. Comparing the operators $\eta_{\sigma}$ and ${\cal X}_{\sigma}$ for the quasi- and pseudo-Hermitian Hamiltonians discussed in Section~3, and noting that according to Eqs.~(\ref{gen-P}), (\ref{gen-PT}), and (\ref{gen-CPT}), ${\cal P}$ is an example of $\eta_{\sigma}$ and ${\cal PT}$ and ${\cal CPT}$ are examples of ${\cal X}_{\sigma}$, we introduce 
	\bea
	{\cal P}&:=&\sum_{\nu_0}\sum_{a=1}^{d_{\nu_0}}
		 (-1)^{\nu_0} |\phi_{\nu_0},a\kt\br\phi_{\nu_0},a|+
		\sum_{\nu}\sum_{a=1}^{d_{\nu}}
		(|\phi_{\nu_+},a\kt\br\phi_{\nu_-},a|+|\phi_{\nu_-},a\kt\br\phi_{\nu_+},a|),
	\label{gen-P-ph}\\
	{\cal T}&:=&\sum_{\nu_0} (-1)^{\nu_0}\sum_{a=1}^{d_{\nu_0}}
		 |\phi_{\nu_0},a\kt\star\br\phi_{\nu_0},a|+
		\sum_{\nu}\sum_{a=1}^{d_{\nu}}
		(|\phi_{\nu_+},a\kt\star\br\phi_{\nu_-},a|+|\phi_{\nu_-},a\kt\star\br\phi_{\nu_+},a|),
	\label{gen-T-ph}\\
	{\cal C}&:=&\sum_{\nu_0} (-1)^{\nu_0}\sum_{a=1}^{d_{\nu_0}}
		 |\psi_{\nu_0},a\kt\br\phi_{\nu_0},a|+
		\sum_{\nu}\sum_{a=1}^{d_{\nu}}
		(|\psi_{\nu_+},a\kt\br\phi_{\nu_-},a|+|\psi_{\nu_-},a\kt\br\phi_{\nu_+},a|),
	\label{gen-C-ph}
	\eea
where we have used the conventions of Sections~3 and~4. 

Again we can check that Eqs.~(\ref{nilp}) and (\ref{C==}) hold. Furthermore, 
	\bea
	{\cal PT}&=&\sum_{\nu_0}\sum_{a=1}^{d_{\nu_0}}
		|\psi_{\nu_0},a\kt\star\br\phi_{\nu_0},a|+
		\sum_{\nu}\sum_{a=1}^{d_{\nu}}
		(|\psi_{\nu_+},a\kt\star\br\phi_{\nu_-},a|+|\psi_{\nu_-},a\kt\star\br\phi_{\nu_+},a|),
	\label{gen-PT-ph}\\
	{\cal CPT}&=&\sum_{\nu_0}(-1)^{\nu_0}\sum_{a=1}^{d_{\nu_0}}
		|\psi_{\nu_0},a\kt\star\br\phi_{\nu_0},a|+
		\sum_{\nu}\sum_{a=1}^{d_{\nu}}
		(|\psi_{\nu_+},a\kt\star\br\phi_{\nu_-},a|+|\psi_{\nu_-},a\kt\star\br\phi_{\nu_+},a|).\nn\\
	&&
	\label{gen-CPT-ph}
	\eea
In view of Eqs.~(\ref{gen-C-ph}), (\ref{gen-PT-ph}) and (\ref{gen-CPT-ph}), Proposition~4, and
the construction given in the proof of Corollary~1, we can check that the operators ${\cal C}$, 
${\cal PT}$, and ${\cal CPT}$ are involutions of the Hilbert space commuting with the Hamiltonian $H$. Therefore, we have the following generalization of Theorem~1.
	\begin{itemize}	
	\item[] {\bf Theorem~2:} Every diagonalizable pseudo-Hermitian Hamiltonian $H$ with a discrete spectrum is invariant under the action of ${\cal C}$, ${\cal PT}$, and ${\cal CPT}$. These operators which are involutions of the Hilbert space generate broken symmetries $H$.
	\end{itemize}

We wish to conclude this section by pointing out that the operators ${\cal P}, {\cal T}$, and 
${\cal C}$ are determined by a complete biorthonormal system associated with the Hamiltonian $H$. As the latter is unique only up to invertible symmetries of $H$, so are these operators.

\section{Conclusion}

In this article, we discussed certain properties of pseudo-Hermitian operators and demonstrated their application in understanding the mathematical origin and exploring generalizations of the findings of Bender, Brody, and Jones \cite{bbj}. In particular, for arbitrary diagonalizable pseudo-Hermitian Hamiltonians with a discrete spectrum, we introduced generalized parity, time-reversal, and charge-conjugation operators that coincide with the ordinary parity, time-reversal, and charge-conjugation for the $PT$-symmetric Hamiltonians (\ref{h}). The generalized parity-time-reversal and charge conjugation operators are examples of generators of a set of  generic symmetries of every diagonalizable pseudo-Hermitian Hamiltonians having a discrete spectrum. A common property of these symmetries is that they are generated by involutions. The generalized parity and time-reversal operators are however involutions only under certain conditions.

\section*{Acknowledgments}

This project was supported by the Young Researcher Award Program (GEBIP) of the Turkish Academy of Sciences.

\ed